\documentstyle[12pt]{article}
\def\be{\begin{equation}}
\def\ee{\end{equation}}
\def\lb{\label}
\def\a{{\bf a}}
\def\b{{\bf b}}
\def\c{{\bf c}}
\def\d{{\bf d}}

\begin{document}

\title{Quantum Group Gauge Theories and Covariant Quantum
Algebras\thanks{Dubna Preprint JINR E2-93-54
(1993) (revised version of the
Wroclaw University Preprint ITP UWr 827/93).}.}

\author{A.P. Isaev\thanks{e-mail address: isaevap@theor.jinrc.dubna.su}\\
\it Laboratory of Theoretical Physics, JINR, Dubna, \\
\it SU-101 000 Moscow, Russia   \\ \\
\rm Z. Popowicz\thanks{e-mail address: ziemek@plwruw11.bitnet} \\
\it Institute of Theoretical Physics,University of Wroclaw \\
\it Pl. Maxa Borna 9, 50-204 Wroclaw, Poland}

\date{}

\maketitle

\begin{abstract}
The algebraic formulation of the quantum
group gauge models in the framework of the $R$-matrix approach
to the theory of quantum groups is given.
We consider gauge groups taking values in the quantum groups
and noncommutative
gauge fields transformed as comodules under the coaction
of the gauge quantum group  $ G_{q}$.
Using this approach we construct
the quantum deformations of the
topological Chern-Simons models, non-abelian gauge theories and
the Einstein gravity. The noncommutative fields in these models
generate $ G_{q}$-covariant quantum algebras.
\end{abstract}

\newpage

\section{Introduction}
\setcounter{equation}0

The quantum deformations
of the non-abelian gauge field theories have
been discussed recently in many papers \cite{AV}-\cite{Cas}.
The basic idea of some of these papers
\cite{IP},\cite{WZh}-\cite{Wat} is to consider the gauge
transformation as a coadjoint action of a quantum group
${\cal G}_{q}$ on the comodule
$F_{ij}$
\begin{equation}
F_{il}
\longrightarrow
\left( T_{ij}  T^{-1}_{kl} \otimes F_{jk} \right) \ , \;\;
\bar{Z} \longrightarrow \ {\cal G}_{q} \otimes
\bar{Z} \ .
\label{1.1}
\end{equation}
where the $N \times N$ matrix elements $\{ T_{ij} \}$
are the generators of the quantum group ${\cal G}_{q}$
while the elements
$F_{il}$ are the generators of some
associative noncommutative algebra with unity $\bar{Z}$.
Here, the quantum group ${\cal G}_{q}$ and the comodule
$F_{ij}$ are interpreted as the
noncommutative analogs of the gauge group and
a curvature (gauge field strength) 2-form, respectively.
In the framework of this interpretation one can try to rewrite
the comodule $F_{il}$ as
$F^{\mu\nu}_{il}dx_{\mu}\wedge dx_{\nu}$ using the basis of
1-forms $dx_{\mu}$ where the
space-time coordinates
$\{ x_{\mu} \}$
could be treated as the usual coordinates of the classical
(Minkowski) space (see \cite{AV}-\cite{BM},
\cite{WZh}-\cite{Cas}) or as noncommutative coordinates
(see \cite{BMa}).

It is tempting, according to the classical case,
to realize the field strength 2-forms $F_{ij}$ by the
square of the covariant differential
$\nabla_{ik} = \delta_{ik} d - A_{ik}$
\be
\lb{1.2}
F_{ij} \equiv - \nabla_{ik} \nabla_{kj} = d(A_{ij}) -
A_{ik} \wedge A_{kj} \; .
\ee
where the gauge potential 1-forms $A_{ik}$
are certain noncommutative objects.
In our paper \cite{IP} (see also \cite{Hir},\cite{Ber},\cite{Cas}),
we have assumed that it is possible to represent $A_{ij}$ as
$A^{\mu}_{ij}dx_{\mu}$ and to define the differential $d$ as
$d=dx_{\mu}\partial^{\mu}$
where $\partial^{\mu}$
are the usual derivatives over the classical space - time coordinates.
Then, for the Yang-Mills curvature 2-form
$F_{ij}$ we obtained the expression
\begin{equation}
\begin{array}{rl}
F^{\mu\nu}_{ij}dx_{\mu}dx_{\nu} \; \equiv &
-\frac{1}{2} \left[
\nabla^{\mu}, \nabla^{\nu} \right]_{ij} dx_{\mu}dx_{\nu} =
 \cr\cr = \frac{1}{2} \{ \left( \partial^{\mu}A^{I\nu} -
 \partial^{\nu}A^{I\mu} \right)
\otimes \sigma^{I}_{ij} \; + &
 (A^{I\mu} A^{J\nu} - A^{I\nu} A^{J\mu})
\otimes \sigma^{I}_{ik} \sigma^{J}_{kj} \} dx_{\mu}dx_{\nu} \; ,
\end{array}
\label{1.3}
\end{equation}
where
$\nabla^{\mu}_{ik} = \delta_{ik} \partial^{\mu}
- A^{I\mu}(x)\sigma^{I}_{ik}$,
$\{ \sigma^{I}_{ij} \}$
is a basis of the classical gauge Lie algebra represented
by $N \times N$ matrices
and we imply the wedge product
in the multiplication of the differential forms
(we will omit $\wedge$ in all formulas below).

In all formulations \cite{AV}-\cite{Cas} of the
noncommutative gauge field theories
important opened problems were
to give the explicit representation of the
noncommutative fields $A_{ij}, \; F_{ij}, \dots $
by some operator functions of $x_{\mu}$ and
to identify the algebras generated by these fields
(see, however, some proposals in \cite{WZh},\cite{Wat}-\cite{Cas}).
In this paper, we would like to continue the investigations
of the papers \cite{AV}-\cite{Cas} in order to fill this gap.
We will consider
the special case when the quantum group ${\cal G}_{q}$ is
$GL_{q}(N)$.
We present, for this case,
the explicit definition of the
noncommutative gauge fields $A_{ij}, \; F_{ij}, \; \dots $
as the generators of the $GL_{q}(N)$-covariant
quantum algebras \cite{IPy}
which have  appeared in the
bicovariant differential calculus on the $GL_{q}(N)$
\cite{Wor,Zum}.  In particular, we
obtain that the 1-forms $A_{ij}$ (2-forms $F_{ij}$) obey the $q$-deformed
anticommutaion (commutation) relations typical of the Cartan 1-forms
(invariant vector fields)
defined on the $GL_{q}(N)$-group \cite{IPy,Zum} while
bosonic (fermionic or veilbein 1-form) matter fields $e_{i}$
commute as coordinates of the quantum hyperplane.
Finally, the closed algebra of the covariant noncommutative fields
$\{ e_{i}, \; A_{ij}, \; F_{ij} \}$ is given by the following
structure relations:
\begin{equation}
\label{1.3a}
\begin{array}{l}
R_{12}e_{1}e_{2} = c e_{2}e_{1} \; , \;\;
R_{12}A_{1}R_{21}A_{2} + A_{2}R_{12}A_{1}R^{-1}_{12} =0 \; , \\ \\
e_{1}A_{2} = (\pm) R_{21}A_{2}R_{12}e_{1} \; , \;\;
R_{12}F_{1}R_{21}F_{2} = F_{2}R_{12}F_{1}R_{21}  \; , \\ \\
e_{1}F_{2} = R_{21}F_{2}R_{12}e_{1} \; , \;\;
R_{21}A_{2}R_{12}F_{1} = F_{1}R_{21}A_{2}R_{12}  \; .
\end{array}
\end{equation}
Here, the indices
$1,2$ enumerate the matrix (for $A,F$)
and vector (for $e$) spaces,
$R_{12}$ is the $R$-matrix for the $GL_{q}(N)$,
$c=q, \; (\pm)=+1$ for the scalar bosons,
$c=-1/q, \; (\pm)=+1$ for the fermions,
$c=-1/q, \; (\pm)=-1$ for the veilbein 1-forms
and $q$ is a parameter of deformation.
In the formulas (\ref{1.1}), (\ref{1.2}) and (\ref{1.3a})
we do not specify the space-time and we do not use
the explicit expansion of the differential forms $e$, $A$ and $F$
in the basis of the 1-forms $dx_{\mu}$.
As we will see below it is very difficult to use the classical
space-time for the algebraic construction presented in this paper.
Note, however, that the naive formulas (\ref{1.3})
typical for the classical space-time
are more attractive from the point of view of
physical applications than their
algebraic analog (\ref{1.2}) considered here and
appropriate for the formulation of the quantum group gauge models
essentially based on the
abstract theory of quantum groups \cite{Dri,Man}.

In this paper, we use the R-matrix formalism \cite{FRT}
which is extremely convenient for the formulation of the
bicovariant differential calculus on the quantum groups.
In the next section we consider
the abstract algebraic construction of the noncommutative gauge
theories. Particularly, we build up the noncommutative
analogs of the Chern-Simons Lagrangians.
In Sect.3 we discuss
unsolved problems such as the problem of the
explicit realization
of the quantum group generators $T_{ij}$
via the operator functions of the space-time coordinates $x_{\mu}$.

\section{$Z_{2}\;$--Graded Extension of $\ GL_{q}(N) \ $
and Quantum Group Gauge Theories.}
\setcounter{equation}0

Let us consider a
$Z_{2}$--graded extension of the
algebra of functions on the
linear quantum group with $GL_{q}(N)$ generators
$\{T_{ij} \}$ and additional new generators $\{ (dT)_{kl}\}$
$ (i,j,k,l=1,2,....,N)$  satisfying the following
commutation relations (see \cite{IPy,Zum,ArV} and references therein):
\begin{eqnarray}
R_{12}T_{1}T_{2} & = & T_{2}T_{1}R_{12} \; ,
\lb{2.1a} \\
R_{12}(dT)_{1} T_{2} & = & T_{2}(dT)_{1}R_{21}^{-1} \; ,
\lb{2.1b}
 \\
R_{12}(dT)_{1} (dT)_{2} & = & - (dT)_{2} (dT)_{1}R_{21}^{-1} \; .
\lb{2.1c}
\end{eqnarray}
where $R_{12}=R^{i_{1},i_{2}}_{j_{1},j_{2}}$ is the
$GL_{q}(N)$ $R$--matrix satisfying the Hecke relation
\be
R_{12}=R^{-1}_{21} + \left( q - q^{-1} \right)P_{12},
\lb{2.2}
\ee
$P_{12}$ is the permutation matrix.
Here and below we use the notation of Ref.\cite{FRT}.
The new generators $(dT)_{ij}$ can be considered as
the differential 1-forms on the quantum group
\cite{IPy}-\cite{Zum}.
It appears that the $Z_{2}$-algebra
with the commutation relations (\ref{2.1a})-(\ref{2.1c})
(we denote this
$Z_{2}$-extension of $Fun(GL_{q}(N))$ by
${\cal G}$ and call it the $Z_{2}$-graded quantum group
in what follows) is simply
the Hopf algebra. Indeed, the comultiplication
$\Delta$, the counit $\epsilon$   and the
antipode $S$ are defined by
\be
\begin{array}{c}
\Delta (T) = T \otimes T \ , \;\;
\epsilon(T) = 1 \  , \;\;
{\cal S} (T) = T^{-1} \ , \cr
\Delta(dT) = dT \otimes T + T \otimes dT
\ , \;\; \epsilon(dT) = 0 \  ,\;\;
 {\cal S} (dT) = - T^{-1} dT T^{-1} ,
\end{array}
\lb{2.3}
\ee
and one can check the following axioms:
\be
\lb{2.4}
\begin{array}{c}
(id \otimes \Delta) \Delta ({\cal G}) =
(\Delta \otimes id) \Delta ({\cal G}) \; ,  \cr
m(\epsilon \otimes id) \Delta({\cal G}) =
m(id \otimes \epsilon) \Delta({\cal G}) = {\cal G}\; ,  \cr
m({\cal S} \otimes id) \Delta({\cal G}) =
m(id \otimes {\cal S}) \Delta({\cal G}) =1\epsilon({\cal G}) \; .
\end{array}
\ee

For our next purposes we introduce also the
$Z_{2}$--graded Zamolodchikov
algebra (denoted by $Z$) generated by the operators
$\{ e_{i},(de)_{i} \}$
$  \ (i=1,2,...N)$
 with the following commutation relations:
\be
\begin{array}{l}
R_{12} e_{1} e_{2} = ce_{2}e_{1} \; , \;\;
(\pm)cR_{12}(de)_{1}e_{2} = e_{2} (de)_{1} , \cr\cr
R_{12}(de)_{1}(de)_{2} = - {1\over c} (de)_{2}(de)_{1} ,
\end{array}
\lb{2.5}
\ee
One can recognize in these
relations (for $(\pm)=+1$) the Wess-Zumino formulas of the covariant
differential calculus on  the bosonic ($c=q$)
and fermionic ($c=-1/q$) quantum hyperplanes \cite{Man,FRT,WZ} where
$ e_{i} $ are the coordinates of the quantum hyperplane while
$(de)_{i}$ are the associated differentials.
The choice $(\pm)=-1$ corresponds to the case when $e_{i}$ are
bosonic $(c=-1/q)$ and fermionic $(c=q)$ veilbein 1-forms.

Let us introduce the left-coaction
$g_{l}$   of the $Z_{2}$-graded
quantum group   ${\cal G}$ (\ref{2.1a})-(\ref{2.1c})
on the generators of the algebra $Z$ by
virtue of the following homomorphism:
\be
\lb{2.8aa}
e_{i} \stackrel{g_{l}}{\longrightarrow} \
\widetilde{e}_{i} = T_{ij} \otimes e_{j},
\ee
\be
\lb{2.8bb}
(de)_{i} \stackrel{g_{l}}{\longrightarrow}
(\widetilde{de})_{i} = (dT)_{ij} \otimes
e_{j} + T_{ij} \otimes (de)_{j} ,
\ee
or in the equivalent form
\be
\lb{2.8cc}
\pmatrix{e \cr de } \stackrel{g_{l}}{\longrightarrow}
\pmatrix{T & 0 \cr dT &T}\otimes \pmatrix{e \cr de} \ .
\ee
The algebra $Z = \{ e, de \}$ becomes now a
left-comodule of ${\cal G}$     with respect
to the coaction (\ref{2.8aa})-(\ref{2.8cc}).
One can verify that all axioms for the comodule are
fulfilled. For example,
we have the following identity:
\be
\lb{2.9}
(\Delta \otimes id)g_{l} ( Z )= (id \otimes g_{l})g_{l}( Z) \; .
\ee
The algebra
$Z$ with the generators (\ref{2.5}) has the following expansion
$Z = \bigoplus\limits_{n=0}\Omega^{n}(Z)$
where $\Omega^{n}(Z)$ denotes the subspace
of the differential n-forms.
Notice that there exists a similar expansion for the
$Z_{2}$-graded
quantum group  (\ref{2.1a})-(\ref{2.1c}):
${\cal G} = \bigoplus\limits_{n=0} \Omega^{n}({\cal G})$.

We have already mentioned that the generators
$(de)_{i}$ and $(dT)_{ij}$
could be considered as differentials
of the variables $e_{i}$ and $T_{ij}$.
One can show that it is possible to
extend the action of the
differential $d$ uniquely to the whole algebra
$ {\cal G} \otimes Z$ in such a way that $d$
obeys the graded Leibnitz rule (e.g. $d( g \otimes Z)=
d( g ) \otimes Z + (-1)^{k} g \otimes d(Z)$,
where $g \in \Omega^{k}({\cal G})$)
and $d^{2} = 0$.

We have introduced the left-coaction
$g_{l}$ (\ref{2.8aa})-(\ref{2.8cc}) because
we would like to interpret it
as a quantum group gauge transformation where the matrix $T_{ij}$
is a noncommutative analog of
a gauge group element and $e_{i}$ are
analogs of the components of the bosonic
or the fermionic (veilbein 1-form) matter fields.
Now the operations
$id \otimes g_{l}$ or $\Delta \otimes id$ in (\ref{2.9})
can be interpreted as the second quantum group gauge transformations.

It is natural to consider the additional term $(dT)_{ij} \otimes e_{j}$
presented in (\ref{2.8bb})
as the effect of the noncovariance of the covector
$(de)_{i}$ under the gauge rotation
(\ref{2.8aa}). The analogous situation
occurs in the
classical gauge theory where derivatives of matter fields are
transformed in the noncovariant way.
Usually, in the classical case, we introduce additional
compensating gauge fields which recover the covariance.
We would like to repeat this trick
in the noncommutative case and
assume that the algebra $Z$ can be extended to
$\bar{Z}$ by adding new elements $A_{ij}$.
We also assume that the operator $d$
can be extended (as a differential) onto the
whole algebra $\bar{Z}$ and hence again
this algebra is decomposed as
\be
\lb{2.12}
\bar{Z} = \bigoplus_{n=0} \Omega^{n} (\bar{Z}) .
\ee
In order to perform our construction we postulate,
first, that the
operators $A_{ij}$ belong to the subspace
$\Omega^{1} (\bar{Z})$ and,
second, the operator
$ (\nabla e)_{i} \in
\Omega^{1}(\bar{Z})$ defined as
\be
\lb{2.13}
(\nabla e)_{i} = (de)_{i} - A_{ij} e_{j} ,
\ee
is transformed homogeneously under (\ref{2.8aa})-(\ref{2.8cc})
as the left-comodule
\be
\lb{2.14}
(\nabla e)_{i} \stackrel{g_{l}}{\longrightarrow}
T_{ij} \otimes (\nabla e)_{j} =
T_{ij} \otimes \left( (de)_{j} - A_{jk}e_{k}\right) .
\ee
According to the classical case we interpret
the compensating operator
$A_{ij}$ satisfying (\ref{2.14})
as a quantum deformation of
a gauge potential 1-form or
as a noncommutative
analog of a connection 1-form.
Using (\ref{2.8aa})-(\ref{2.8cc}) and (\ref{2.14}) one can deduce
the noncommutative analog
of the  gauge transformation for this gauge potential as
\be
\lb{2.15}
A_{ik} \stackrel{g_{l}}{\longrightarrow}
\widetilde{A_{ik}} =
T_{ij} T^{-1}_{lk} \otimes A_{jl}
+ dT_{ij} T^{-1}_{jk} \otimes 1 \; , \;\;
\widetilde{A}_{ij} \in {\cal G} \otimes \bar{Z} .
\ee
It is natural to call the noncommutative connection $A$, satisfying the
transformation rule (\ref{2.15}),
the $q$-deformed gauge comodule.
One can justify this terminology
taking into account the identity
\be
\lb{2.16}
(\Delta \otimes id)g_{l}( A )=
(id \otimes g_{l})g_{l}( A )\; .
\ee
Here we stress again that it is possible to
interpret the action
$\Delta \otimes id $
(or $id \otimes g_{l} $) presented in Eq.(\ref{2.16})
as the second noncommutative gauge transformation and it is
exactly what has been assumed in \cite{IP}.
To define the algebra $\bar{Z}$ explicitly we have to deduce
commutation relations of $A_{ij}$
with the generators $e_{i}$ and $(de)_{j}$.
First of all, we note that the components $A_{ij}$
of the $q$-deformed 1-form connection
generate some closed algebra.
In order to find structure relations for this algebra
we remark
that there is a trivial representation for the generators $A_{ij}$,
namely $A = dTT^{-1} \otimes 1$,
 which is gauge equivalent
to the trivial connection $A_{ij}=0 $
corresponding to the "flat
noncommutative geometry". Using this
representation and the relations (\ref{2.1a})-(\ref{2.1c}) we
conclude that the generators $A_{ij}$
have to satisfy the following $q$-deformed
anticommutation relation:
\be
\lb{2.17}
R_{12} A_{1} R_{21}A_{2} + A_{2} R_{12}A_{1}R^{-1}_{12} = 0 .
\ee
Note that the same relations
for the noncommutative gauge fields have been postulated also
in \cite{Wat,Cas}.
Now one can show directly that if
$A_{ij}$ obey (\ref{2.17})
then the gauge transformed operators
$\widetilde{A}_{ij}$ (\ref{2.15}) satisfy (\ref{2.17})
too. Hence, the transformation (\ref{2.15}) is the homomorphism
of the algebra (\ref{2.17}).
In order to find the commutation relations $A_{ij}$ with
the generators $\{ e_{i}, \; (de)_{i} \}$ we postulate that
the coordinates of the covector (\ref{2.13}) commute
in the same way as the components
of 1-forms $(de)_{i}$ (see Eqs.(\ref{2.5}))
$R_{12}(\nabla e)_{1}(\nabla e)_{2} =
- {1\over c} (\nabla e)_{2}(\nabla e)_{1}$.
Using Eq.(\ref{2.17}) one can show that
this relation is equivalent to two covariant relations
\be
\lb{2.19}
e_{1}A_{2} =(\pm) R_{21}A_{2}R_{12}e_{1} \; ,
\ee
\be
\lb{2.20}
(\pm)(de)_{1}A_{2}=-R_{21}A_{2}R_{12}(de)_{1} +
(q-q^{-1})P_{12}A_{2}R_{12}A_{1}e_{1} \; .
\ee
These are unique relations which are
covariant under the gauge
co-transformations (\ref{2.8aa})-(\ref{2.8cc}), (\ref{2.15})
and allow one to push the operators $\{ e_{i}, \; (de)_{i} \}$ through
the operators $A_{kl}$.
The concise covariant form for Eq.(\ref{2.20}) is
\be
(\nabla e)_{1} A_{2} = -(\pm)R_{21}A_{2}R_{12}(\nabla e)_{1} \; .
\lb{2.21}
\ee
Now applying the second covariant derivative $\nabla$ to the
expression (\ref{2.13}) and using the $Z_{2}$-graded Leibnitz rule
we can define the noncommutative analog of the
field strength (curvature) 2-form $F$.
As a result, we have
\be
\lb{2.22}
\nabla(\nabla e) = - \left( d(A) - A^{2}\right) e = - Fe .
\ee
The next action of the covariant
derivative on the formula (\ref{2.22}) yields the
Bianchi identities which can be written in the
classical form $d(F)=[A , \; F]$.
It is clear that the quantum gauge
transformation (\ref{2.15}) for the curvature
$F$ is represented as the coadjoint action
\be
\lb{2.23}
F_{ij}{ \stackrel{g_{ad}}{\longrightarrow}}
\ \widetilde{F}_{ij}
= \left( T_{ik} T^{-1}_{lj} \right) \otimes F_{kl} \; .
\ee
The commutation relations for the operators $dA_{ij}$ and
$F_{ij} = dA_{ij} -A_{ik}A_{kj}$ can be deduced from
the quantum hyperplane condition
\be
\lb{2.24}
R_{12}(Fe)_{1}(Fe)_{2} = c (Fe)_{2}(Fe)_{1} \; .
\ee
Differentiating Eq.(\ref{2.19}) and using (\ref{2.19}), (\ref{2.20})
and (\ref{2.24}) one can derive the relations
\be
\lb{2.25}
e_{1}F_{2} = R_{21}F_{2}R_{12}e_{1} \; , \;\;
P_{12}^{-}F_{1}R_{21}F_{2}P^{+}_{12}=0 \; ,
\ee
where we have introduced the projectors
$P^{\pm}_{12} = (R_{12} \mp q^{\pm 1} P_{12})/(q+q^{-1})$.
The last equality in (\ref{2.25})
is the consequence (see \cite{IPy})
of the $q$-deformed commutation
relations
\be
\lb{2.26}
R_{12}F_{1}R_{21}F_{2} -F_{2}R_{12}F_{1}R_{21} =0
\ee
which are the same for the invariant vector fields defined on
the $GL_{q}(N)$-group (see \cite{IPy}-\cite{Zum}).
Eqs.(\ref{2.26}) are known as the reflection equations \cite{Kul}
and also are the structure relations for the
braided algebras \cite{Mad}. To complete the definition of the
algebra $\bar{Z}$
we present the unique covariant commutation relations for
$F$ and $A$
\be
\lb{2.27}
F_{1}R_{21}A_{2}R_{12} = R_{21}A_{2}R_{12}F_{1} \; .
\ee
The commutation relations (\ref{2.5}), (\ref{2.17}),
(\ref{2.19})-(\ref{2.21}) and (\ref{2.25})-(\ref{2.27})
completely define the algebra $\bar{Z}$.
We stress that this algebra is covariant under the gauge
group coactions (\ref{2.8aa})-(\ref{2.8cc}) and (\ref{2.15}).
We note that
the possible relation
$$
R_{12}(dA)_{1}R_{21}(dA)_{2} =
(dA)_{2}R_{12}(dA)_{1}R_{21}
$$
(noncovariant under the gauge
co-transformations (\ref{2.15}) and (\ref{2.23}))
postulated in \cite{Wat}
is inconsistent with Eqs.(\ref{2.26}) and (\ref{2.27}) and
is not fulfilled in our approach.

Our final aim is to define the
noncommutative Lagrangians which describe
the quantum group gauge theories. To write
down the Lagrangians invariant
under the co-transformations
(\ref{2.8aa})-(\ref{2.8cc}), (\ref{2.15})
we extend $\bar{Z}$ by virtue of introducing
the $Z_{2}$-graded contragradient comodule
$\left( \bar{e}_{i}, d\bar{e}_{j} \right)$
with the following commutation relations:
\be
\lb{2.28}
\begin{array}{l}
\bar{e}_{1}\cdot \bar{e}_{2} R_{12} = c \bar{e}_{2}\bar{e}_{1} \; , \;\;
(d \bar{e})_{1} \bar{e}_{2} =
(\pm)c \bar{e}_{2} (d \bar{e})_{1} R_{21} , \cr\cr
(d \bar{e})_{1} (d\bar{e})_{2} R_{12} = -
{1 \over c} (d\bar{e})_{2} (d\bar{e})_{1} .
\end{array}
\ee
The quantum group gauge transformation of the vector
($\bar{e}_{i}, d\bar{e}_{j}$) is expressed as the
following homomorphism
of the algebra (\ref{2.28}):
\be
\lb{2.29}
( \bar{e}, \; d\bar{e} )
{\stackrel{g_{r}}{\longrightarrow } }
(\bar{e},\; d\bar{e})
\otimes\pmatrix{T^{-1}, & -T^{-1}dT T^{-1} \cr 0 , & T^{-1} } ,
\ee
where the generators
$T_{ij}$ and
$dT_{kl}$ are the same as in Eqs.(\ref{2.1a})-(\ref{2.1c}).
The commutation relations of the contragradient generators
$\{ \bar{e}_{i}, \; d\bar{e}_{j} \}$
with the former generators of $\bar{Z}$ can be found using
covariance of these relations under the gauge co-actions
(\ref{2.8aa})-(\ref{2.8cc}), (\ref{2.15}) and (\ref{2.29}).
For example, one can deduce:
$$
c e_{1}\bar{e}_{2} = \bar{e}_{2} R^{-1}_{21} e_{1} \; , \;\;
(\pm)c (de)_{2} \bar{e}_{1} = \bar{e}_{1} R_{21} (de)_{2} \; ,
$$
$$
A_{1}\bar{e}_{2} = (\pm) \bar{e}_{2} R_{12} A_{1} R_{21} \; , \;\;
F_{1}\bar{e}_{2} =  \bar{e}_{2} R_{12} F_{1} R_{21} \; .
$$

Now we define the invariant Lagrangian for the noncommutative fields
$e_{i}, \; \bar{e}_{i}$ and $A_{ij}$ as
\be
\lb{2.30}
{\cal L} = \bar{e}_{i} \left( de_{i} - A_{ij}e_{j}\right) .
\ee
One can interpret this as the Lagrangian for the
noncommutative version of
various discrete gauge models (see e.g. \cite{FIG}).

In order to write down other quantum group
gauge invariants, it is possible to use
the field strength 2-form $F$ which transforms  as
the adjoint comodule (\ref{2.23}). Indeed, let us consider
the following Lagrangians:
\be
\lb{2.31}
L^{(k)}_{top}=Tr_{q}(F^{k}) = Tr(DF^{k})=D_{ij}F_{jj_{1}}F_{j_{1}j_{2}}
\dots F_{j_{k-1}i}\; ,
\ee
where we introduce the
notion of the $q$-deformed trace
\cite{IP,FRT,Zum,IM} defined by the matrix
$D_{ij}$. For the $GL_{q}(N)$ case
this matrix is the diagonal matrix
$D_{ij} = q^{2i}\delta_{ij}$.
Using the wellknown feature of the
$q$-trace
$Tr_{q}(TET^{-1})=Tr_{q}(E)$ where
$ [T_{ij}, \; E_{kl} ] =0$,
one can obtain that the expressions
(\ref{2.31}) are invariant under the gauge
coaction (\ref{2.23}).
Moreover, one can show that
$L^{(k)}_{top}$ is a closed $2k$-form and
$L^{(k)}_{top} = d(L^{(k)}_{CS})$
where we have introduced the $(2k-1)$-form
\be
\lb{2.32}
L^{(k)}_{CS}=Tr_{q} \{ A(dA)^{k-1} + \frac{1}{h^{(k)}_{1}} A^{3}(dA)^{k-2} +
\dots + \frac{1}{h^{(k)}_{\dots}} A^{2k-1} \}
\ee
Notice that $L^{(k)}_{CS}$ could be interpreted as the
noncommutative Chern-Simons Lagrangians.
The constants $h^{(k)}_{i}$ in (\ref{2.32}) depend on
the choice of the quantum group ${\cal G}_{q}$ and the parameter of
deformation $q$. For example, in the case
of the $GL_{q}(N)$-group, $L_{CS}^{(2)}$
reproduces the noncommutative analog of the
three-dimensional Chern-Simons term and we obtain
\be
\lb{2.33}
L_{CS}^{(2)} = Tr_{q} \{ AdA - \frac{1}{h^{(2)}_{1}} A^{3} \} \; , \;\;
h^{(2)}_{1} = 1 + \frac{1}{q^{2} + q^{-2}} \; .
\ee

Let us stress that we have not specified the space-time $M$
which can be commutative as well as
non-commutative. There are some arguments (see e.g. Conclusion)
that it is rather difficult to use classical space-time for
the quantum group gauge theories presented here. That is why
we assume that the space-time $M$ is the
$n$-dimensional quantum hyperplane with
coordinates $\{ x_{\mu} \}$.
One can define for such space $M$ the
metric tensor $C_{\mu\nu}$ (e.g. the metric matrix of $SO_{q}(n)$),
the $n$-form ${\cal E}=dx_{1} \cdots dx_{n}$ which is
the analog of the volume element and construct
the dual isomorphism $"*"$ of the differential forms on $M$ (Hodge map).
Then, we postulate that
there exists a map $\pi^{-1} : \; M \mapsto \bar{Z}$ and
it is possible to expand the differential forms $F_{ij}$
over the basis of the 1-forms
$dx_{\mu}$
\be
F_{ij} = F^{\mu\nu}_{ij}dx_{\mu}dx_{\nu} \; , \;\;
*F_{ij} =dx_{\mu_{1}} \dots dx_{\mu_{n-2}}
{{\cal E}_{q}^{\mu_{1} \dots \mu_{n-2}} }_{\mu \nu} F_{ij}^{\mu\nu} \; .
\lb{2.34}
\ee
The coefficients $F^{\mu\nu}_{ij}, \dots$ of
the differential forms are operator functions of $\{ x_{\mu}\}$.
The quantum group gauge invariant Lagrangian for
the q-deformed Yang-Mills field theory ($n=4$)
can be represented, following the line proposed in \cite{IP},\cite{Hir} as
\be
\lb{2.35}
{\cal L} = Tr_{q} \left( F * F \right) \sim
 D_{ij}F^{\mu\nu}_{jk} {\cal E} F_{ki,\mu\nu} \; .
\ee

Another attractive possibility is the choice of the space-time $M$
isomorphic to the space of the quantum group $GL_{q}(N)$. In this
case it is tempting to explore monopole-instanton type gauge potential
1-forms
$([\tilde{A}, T ] = 0, \;
\tilde{A}(z)dT  = dT\tilde{A}(q^{2}z))$
$$
A_{ij}=dT_{ik}\tilde{A}_{kl}(z)T_{lj}^{-1} =
dT_{ik}T_{lj}^{-1}\tilde{A}_{kl}(z) \; ,
$$
where $z=det_{q}T$
and $
\tilde{A}_{1}(q^{2}z)R_{12}^{-1}\tilde{A}_{2}(z)R_{21}^{-1} -
R_{12}^{-1}\tilde{A}_{2}(q^{2}z)R_{21}^{-1}\tilde{A}_{1}(z) = 0 \; .  $

Along the same way as above one can formulate the
noncommutative version of the
Einstein theory of gravity
(see another approach presented in \cite{Cast}).
For this purpose
we have to take the
underlying Zamolodchikov algebras (\ref{2.5}), (\ref{2.28}) in the
form with $c= -1/q, \; (\pm)=-1$ and interpret
$\{ e_{i}, \; \bar{e}_{i}=e_{j}C_{ji} \}$      as the
noncommutative analogs of the vielbein 1-forms.
Then, we assume the expansion of $F_{ij}$ in the form:
$F_{1}=\bar{e}_{2}F_{1,2}e_{2}$.
In this case, the Lagrangian ${\cal L}= e_{i_{1}} \dots e_{i_{n}}
{\cal E}_{q}^{i_{1} \dots i_{n}}F$,
where $F=Tr_{q1}Tr_{q2}(F_{1,2}P_{12}R_{12})$
is a scalar curvature, describes
the noncommutative Einstein gravity and
the gauge quantum group
${\cal G}_{q}$ must be the quantum
Lorentz group considered in \cite{CW}.
Note that the quantum trace plays
the essential role in our formulation of the noncommutative
gauge theories. At the end of this Section we stress that the
transformations (\ref{2.8aa}), (\ref{2.8bb}) were also discussed in
\cite{Sud}.

\section{Conclusion}
\setcounter{equation}0

Our aim in this paper has been to present the algebraic formulation
of the $q$-deformed non-abelian gauge theories discussed in our
previous paper \cite{IP}. It appears that the noncommutative analogs
of the gauge and matter fields generate various covariant quantum
algebras specific for the covariant differential calculi on
the quantum groups and quantum hyperplanes.
After this, the next step must be the
construction of the explicit representations of these covariant
algebras and especially the explicit realization of the maps
\be
\lb{3.0}
M \mapsto \bar{Z} \; {\rm and} \; M \mapsto {\cal G} \; ,
\ee
where $M$ is
a classical (or quantum) space-time. The solution of this problems
is extremely important for the realization of the noncommutative
fields in terms of the classical fields in order to find
the field theoretical interpretation to the algebraic construction
presented in the previous section.
For example, we stress that
the Lagrangians (\ref{2.30})- (\ref{2.32}) are
in general the noncommutative elements of the algebra $\bar{Z}$ and
to obtain the usual Lagrangians we have to make an additional averaging
over this algebra (see \cite{IP}).
However, it is worth notice that
the Lagrangian (\ref{2.30})
is the central element for the subalgebra of
$\bar{Z}$ generated by the operators $A_{ij}$, $F_{ij}$, $e_{i}$ and
$\bar{e}_{i}$ while
(\ref{2.31}) is the central element for the
subalgebra with the generators $A_{ij}$ and $F_{ij}$. Unfortunately,
these Lagrangians are not the central elements
for the whole algebra $\bar{Z}$.

To conclude this paper, we illustrate the
problem (related to the explicit construction
of the maps (\ref{3.0}))
by considering the map $M \mapsto GL_{q}(2)$
where $M$
is a classical space-time.
Let us explore the following
obvious realization of $T_{ij}(x)$ for the $GL_{q}(2)$-group
\be
T_{ij}(x) =
\pmatrix{\alpha(x,z)\a + \bar{\delta}(x,z,\b\c)(1 / \d ) &
\beta(x,z)\b \cr
\gamma(x,z) \c & \delta(x,z)\d + (1/ \a ) \bar{\alpha}(x,z,\b\c) } \; ,
\lb{3.2}
\ee
where $T_{ij}=\pmatrix{\a & \b \cr \c & \d }$ is the standard notation
for the $GL_{q}(2)$ generators \cite{FRT} ($\a$ and $\d$ are invertible
generators),
$\bar{\alpha}=\sum_{i=0}
\bar{\alpha}_{i}(\b\c)^{i}\; , \;
\bar{\delta} = \sum_{i=0}
\bar{\delta}_{i}(\b\c)^{i}
$, $z=det_{q}(T)=\a\d-q\b\c$ is the
central element for $GL_{q}(2)$
and $\alpha , \; \beta , \; \gamma , \;
\delta , \; \bar{\alpha}_{i}$ and $\bar{\delta}_{i}$
are classical functions of
$x_{\mu}$ and $z$ which could be
considered as the parameters of the gauge
quantum group.
One can
prove that $C(x)=det_{q}(T(x))$
is the central element for $GL_{q}(2)$
and the operators (\ref{3.2}) commute as in (\ref{2.1a}) only if
\be
\lb{3.3}
\begin{array}{l}
(qW + \kappa_{1})z^{2} +
(\bar{\kappa}_{1}z-q\bar{\kappa}_{0})  = 0 \; , \;\;
q(qW + \kappa_{1}) + \kappa_{2}z + \bar{\kappa}_{2} = 0 \; ,\\ \\
\kappa_{i}z + q\kappa_{i-1} + \bar{\kappa}_{i} = 0 \; \; (i>2)\; ,
\end{array}
\ee
where $\kappa_{i}=\alpha\bar{\alpha}_{i} + \delta\bar{\delta}_{i}$,
$\bar{\kappa}_{i} = \sum_{j=0}^{i}\bar{\delta}_{j}\bar{\alpha}_{i-j}$
and $W=\alpha\delta - \beta\gamma$. From Eqs.(\ref{3.3}) we obtain
$
C(x) \equiv det_{q}(T(x))=\alpha\delta z + \kappa_{0} +
\bar{\kappa}_{0} / z \; .
$
However if we would like to interpret
generators $dT_{ij}(x)$
as the usual differentials of (\ref{3.2}) with respect to the
space-time coordinates $x_{\mu}$ then we obtain the contradiction.
Indeed the $dT_{ij}(x)$ can be written down as
\be
\!\!\! dT_{ij}(x) =  dx_{\mu} \partial^{\mu}
\pmatrix{ \alpha(x)\a +
\bar{\delta}_{i}(x)(\b\c)^{i}\frac{1}{\d} &
 \beta(x)\b \cr
 \gamma(x) \c &
 \delta(x)\d +
 \bar{\alpha}_{i}(x)\frac{1}{\a}(\b\c)^{i} }  .
\lb{3.4}
\ee

{}From here we immediately obtain that (\ref{3.2}) and (\ref{3.4})
do not realize the algebra (\ref{2.1a})-(\ref{2.1c}).
For example, the equation $T_{21}(x) dT_{11}(x) =
q^{-1} dT_{11} T_{21}$
easily deduced from (\ref{3.2}), (\ref{3.4}) obviously contradicts
the commutation
relations (\ref{2.1b}) which
take the following form for the $GL_{q}(2)$ case:
\be
\lb{3.5}
\begin{array}{l}
T_{21}dT_{11} = q dT_{11}T_{21} \ , \;
T_{i2}dT_{1i} = q dT_{1i}T_{i2} \ , \;
T_{ij}dT_{ij} = q^{2} dT_{ij}T_{ij} \ , \\ \\
\!\!\!\!\!\!\! \left[ T_{21}, dT_{12} \right] =
(q-\frac{1}{q})dT_{11}T_{22} \ , \;
\left[ T_{22}, dT_{11} \right] = 0  , \;
d(R_{12}T_{1}T_{2}) = d(T_{2}T_{1}R_{12}) .
\end{array}
\ee
This example shows
that the obvious map $M \mapsto {\cal G}$
(\ref{3.2}) is inconvinient for the construction
of the quantum group gauge theories discussed in the
previous section. Moreover, it seems very difficult to
construct the map (\ref{3.0}) with $M$ which is the classical
space-time.
The realization of the appropriate maps (\ref{3.0})
for the quantum space-time $M$
is in progress now.

\section*{Acknowledgments} The authors would like to thank
I.Ya.Aref'eva, Sh.Majid, P.N.Pyatov and V.N.Tolstoy for
the helpful discussions,
constructive criticism and interest in this paper.

This work was supported in part by the grant KBN No.201049101.

\end{document}